\documentclass[epj,referee]{svjour}
\usepackage{amsmath}
\usepackage{latexsym}
\usepackage[dvips]{graphicx}
\usepackage[dvips]{epsfig}
\newcommand{\beq}{\begin{equation}}
\newcommand{\eeq}{\end{equation}}
\newcommand{\ds}{\displaystyle}
\newcommand{\dd}{\textrm d}

\newcommand{\dgr}{ {\,}^{\circ} \mbox{C}}
\newcommand{\un}[1]{\ensuremath{\unskip\,\mathrm{#1}}}

\begin{document}

\title{Solid-Supported Lipid Multilayers~: Structure Factor and Fluctuations}

\author{Doru Constantin\thanks{\email{dconsta@gwdg.de}} \and
Ulrike Mennicke\thanks{\email{ulrike.mennicke@gmx.net}} \and
Chenghao Li\thanks{\email{cli@uni-goettingen.de}} \and Tim
Salditt\thanks{\email{tsaldit@gwdg.de}}}

\institute{Institut f\"{u}r R\"{o}ntgenphysik, Geiststra{\ss}e 11,
37073 G\"{o}ttingen, Germany.}
\date{Received: date / Revised version: date}
% The correct dates will be entered by Springer
%
\abstract{
We present a theoretical description of the thermal fluctuations in a solid-supported
stack of lipid bilayers, for the case of vanishing surface tension $\gamma = 0$
and in the framework of continuous smectic elasticity. The model
is successfully used to model the reflectivity profile of a thin
(16 bilayers) DMPC sample under applied osmotic pressure and the
diffuse scattering from a thick (800 bilayers) stack. We compare
our model to previously existing theories.
 % end of PACS codes
} %end of abstract
\maketitle
\section{Introduction}
\label{sec:intro}

Lipid bilayer systems have been extensively studied using X-ray techniques, including small-angle
scattering and reflectivity measurements \cite{katsaras00}. Among the various experimental configurations
employed, the case of solid-support\-ed stacks of bilayers is very important from a practical point of
view, as these samples are amenable to pow\-er\-ful, interface-sensitive scattering techniques
(spec\-u\-lar and non\-spec\-u\-lar scattering, grazing incidence difraction) avoiding the ambiguities
associated with powder averaging \cite{salditt02,lyatskaya00}.

Besides the fundamental physical problems they raise, such as the
molecular forces responsible for the inter-bi\-lay\-er interaction
potential (steric, electrostatic, van der Waals etc.); the
influence of static defects; the loss of long-range order in
low-dimensional systems and the role of the fluctuations, these
systems can also serve as testing ground for the interaction of
the cell membrane with membrane active molecules, such as antimicrobial peptides
\cite{bechinger97}. It is then of paramount importance to understand the different
parameters influencing the scattering signal in order to achieve a comprehension
of the spectra before the (sometimes subtle) effect of the included molecules can
be confidently assessed.

One of the most characteristic hallmarks of lamellar systems (exhibiting one-dimensional order) is the
Landau-Peierls effect, whereby the long-range order is destroyed by thermal fluctuations. Ever since the
seminal paper of Caill\'{e} \cite{caille72}, this phenomenon has been studied in great detail, first in the
bulk \cite{alsnielsen80,safinya86,safinya89,nallet93,zhang94,nagle96,pabst00} and subsequently in thin
films \cite{holyst91,shalaginov93,lei95,mol96}, mainly under the influence of extensive experimental
studies on freely-suspended films of thermotropic smectics (see \cite{oswald02} for a review). Recently,
theoretical models were developed for the study of fluctuations in smectic films on solid substrates
\cite{deboer99,romanov02}, taking into account both the surface tension at the free surface and the
boundary conditions imposed by the substrate.

In this paper we describe the thermal fluctuations in a solid-supported
stack of lipid bilayers, concentrating on the (experimentally relevant) case of
vanishing surface tension at the top of the stack. The influence of the substrate
is only considered inasmuch as it limits the fluctuations of the bilayers,
neglecting direct interactions (\textit{e. g.} van der Waals, electrostatic) which
can become important in the case of very thin films.

The paper is structured as follows~: in section \ref{sec:elast} we
discuss the positional fluctuations of the layers in a
solid-supported stack in the framework of continuous smectic
elasticity, taking however into account the discrete nature of the
system by the limitation of the number of modes along the normal
direction. In contrast with (free-standing or supported) films of
thermotropic smectics, for fully hydrated lipid multilayers the
surface tension at the free surface vanishes. This greatly
simplifies the theoretical treatment of the fluctuations. We
conclude this section by a discussion of the in-plane variation of
the correlation function.

Section \ref{sec:scattering} starts with a discussion of the
specular structure factor $S(q_z)$. We then consider the influence
on $S$ of the coverage rate, which can vary due to the preparation
technique or to partial dewetting upon hydration. We compare
our model with experimental data on the specular scattering
and then with an estimate of the in-plane correlation
function obtained from the diffuse scattering signal.

\section{Smectic elasticity}
\label{sec:elast}

We consider in the following a solid-supported sample with
periodicity $d$, of thickness $L=Nd$ and extending over a surface
$S$ in the plane of the layers. We take the origin of the $z$ axis
on the substrate, so that $z=L$ gives the position of the free
surface. The in-plane position is denoted by ${\mathbf
r}_{\bot}=(x,y)$.

\subsection{Model and fluctuations}
\label{subsec:modelfluct}

The simplest description of smectic elasticity is provided by the
continuous smectic hamiltonian~:
 \beq \label{contsmecham}
\begin{split}
F &= \frac{1}{2} \int_{V} \dd {\mathbf r} \left [ B \left (
\frac{\partial u({\mathbf r}_{\bot},z)}{\partial z} \right ) ^2 +
K \left ( \Delta _{\bot} u({\mathbf r}_{\bot},z) \right ) ^2
\right ] \\
 &+ \frac{\gamma}{2} \int_{S} \dd {\mathbf r}_{\bot}
\left ( \nabla _{\bot} u({\mathbf r}_{\bot},L) \right ) ^2 \, .
\end{split}
\eeq

Following the treatment of Poniewierski and Ho{\l}yst
\cite{poniewierski93}, we shall decompose the deformation over
independent modes; first, we take the Fourier transform of
$u({\mathbf r}_{\bot},z)$ in the plane of the bilayer~:
\beq
u({\mathbf r}_{\bot},z) = \frac{1}{\sqrt{S}} \sum_{{\mathbf
q}_{\bot}} \exp (-i {\mathbf q}_{\bot} {\mathbf r}_{\bot})
u({\mathbf q}_{\bot},z) \, , \label{tfqperp}
\eeq
The free energy (\ref{contsmecham}) can now be written as the sum
$F=\sum_{{\mathbf q}_{\bot}} F_{{\mathbf q}_{\bot}}$, with~:
\beq \label{fq}
\begin{split}
F_{{\mathbf q}_{\bot}} &= \frac{1}{2} \int_0^L \dd z \left [ B
\left | \frac{\partial u({\mathbf q}_{\bot},z)}{\partial z} \right
| ^2 + K {\mathbf q}_{\bot}^4 \left | u({\mathbf q}_{\bot},z)
\right | ^2 \right ] \\
&+ \frac{\gamma}{2} {\mathbf q}_{\bot}^2 \left | u({\mathbf
q}_{\bot},L) \right | ^2 \, .
\end{split}
\eeq
The boundary conditions for the $u({\mathbf q}_{\bot},z)$
components are~:
\begin{subequations}
\label{boundary}
\beq u({\mathbf q}_{\bot},0)=0 \eeq \beq \gamma {\mathbf
q}_{\bot}^2 u({\mathbf q}_{\bot},L) +B\left . \frac{\partial
u({\mathbf q}_{\bot},z)}{\partial z} \right | _{z=L} =0 \, .
\eeq
\end{subequations}
The first condition (\ref{boundary}a) simply states that the
fluctuations go to zero at the substrate; the second one
(\ref{boundary}b) is necessary in order to write $F_{{\mathbf
q}_{\bot}}$ in (\ref{fq}) as a quadratic form \cite{shalaginov93}.
Physically, it expresses the continuity across the interface of
the $\sigma _{zz}$ component of the stress tensor.

The correlation function of the fluctuations can be defined as~:
\beq
 C({\mathbf r}_{\bot},z,z')= \left \langle u({\mathbf
r}_{\bot},z) u({\mathbf 0},z') \right \rangle \, , \label{corrfun}
\eeq
\noindent where $ \left \langle \cdot \right \rangle $ denotes the
ensemble average. From the Wiener-Khinchin theorem, its Fourier
transform is~:
\beq
C({\mathbf q}_{\bot},z,z')= \left \langle u({\mathbf
q}_{\bot},z) u(-{\mathbf q}_{\bot},z') \right \rangle \, .
\label{corrfunq}
\eeq
We then expand $u({\mathbf q}_{\bot},z)$ over the orthonormal set
of harmonic functions $f_n(z)$, chosen to fulfill the boundary
conditions (\ref{boundary})~:
\beq
u({\mathbf q}_{\bot},z) = \sum_{n=1}^{N} \delta u_n ({\mathbf
q}_{\bot}) f_n(z) \label{uqz}
\eeq
\noindent where the summation goes from $1$ to $N$, instead of
$\infty$, because only $N$ components are required to describe the
position of the $N$ bilayers (this amounts to restricting the summation
to the first Brillouin zone). Keep in mind, however, that the
index $n$ denotes here a particular {\em deformation mode} and not
an individual {\em bilayer}.

Finally, the free energy can be written as a sum of independent
modes~: $F=\frac{1}{2} \sum_{{\mathbf q}_{\bot}} \sum_{n=1}^{N}
A(n,{\mathbf q}_{\bot}) \left | \delta u_n ({\mathbf q}_{\bot})
\right | ^2$, where the "stiffness" $A(n,{\mathbf q}_{\bot})$
associated to each mode depends on the elastic constants $B$, $K$
and $\gamma$. The equipartition theorem yields~:
\beq
\begin{split}
\left \langle \delta u_n ({\mathbf q}_{\bot}) \delta u_m
(-{\mathbf q}_{\bot}) \right \rangle &= \delta _{mn} \left \langle
\left | \delta u_n ({\mathbf q}_{\bot}) \right | ^2 \right
\rangle \\
 &= \delta _{mn} k_B T / A(n,{\mathbf q}_{\bot}) \, ,
\end{split}
\eeq
\noindent with $\delta _{mn}$ the Kronecker symbol. Plugging
(\ref{uqz}) in (\ref{corrfunq}), one has~:
\beq
C({\mathbf q}_{\bot},z,z')= \sum_{n=1}^{N} f_n(z) f_n(z')
\left \langle \left | \delta u_n ({\mathbf q}_{\bot}) \right | ^2
\right \rangle \, . \label{corrfunq2}
\eeq
We can now Fourier transform back to the real space domain~:
\beq
\label{corrfun2} \begin{split} C(r,z,z')=\frac{1}{2 \pi}
\sum_{n=1}^{N} f_n(z) f_n(z') \\
\int_{0}^{\infty} q_{\bot} \dd q_{\bot} \mathrm{J}_0 (q_{\bot} r)
\left \langle \left | \delta u_n ({\mathbf q}_{\bot}) \right | ^2
\right \rangle \, .
\end{split}
\eeq
\noindent where $r=\left | {\mathbf r}_{\bot}\right |$ is the
in-plane distance (the $\bot$ symbol can safely be omitted).

\subsection{The case of vanishing surface tension ($\gamma = 0$)}
\label{subsec:gammazero}

Until now we have presented the general formalism; in each case,
we must find the orthonormal set of functions $f_n(z)$, which are
selected by the boundary conditions, and then determine the
amplitude of each mode $\left \langle \left | \delta u_n ({\mathbf
q}_{\bot}) \right | ^2 \right \rangle$ (or, equivalently, the
stiffness $A(n,{\mathbf q}_{\bot})$). This is what we shall now do
for the case when $\gamma =0$, which is not only the simplest,
but also the one relevant for systems of fully hydrated membranes \cite{safran99}.

The boundary conditions (\ref{boundary}) are in this case~:
$u({\mathbf q}_{\bot},0)=0$ and $ \left . \frac{\partial
u({\mathbf q}_{\bot},z)}{\partial z} \right | _{z=L} =0$, so the
set of $f_n(z)$ is~:
\beq
\label{fnset} f_n(z) = \sqrt{\frac{2}{L}} \sin \left (
\frac{2n-1}{2} \pi \frac{z}{L} \right ) \, .
\eeq

The amplitudes are given by~:
\beq
\begin{split}
\left \langle \left | \delta u_n ({\mathbf q}_{\bot}) \right | ^2
\right \rangle &= \frac{k_B T}{B \left ( \frac{2n-1}{2}
\frac{\pi}{L} \right )^2+ K {\mathbf q}_{\bot}^4} \\
&= \frac{k_B T}{B} \frac{4L^2}{(2n-1)^2 {\pi}^2} \frac{1}{1+(\xi_n
q_{\bot})^4}
\end{split}
\eeq

\noindent where $\ds \xi_n^2 = \frac{2 L \lambda}{(2n-1) \pi}$,
with $\lambda = \sqrt{K/B}$. We also define the dimensionless
parameter $\ds  \eta = \frac{\pi}{2} \frac{k_B T}{B \lambda d^2}$,
first introduced by Caill\'{e} \cite{caille72}.

Finally, from equation (\ref{corrfun2}) we have~:
\beq \label{corrfunfinal}
\begin{split}
&C(r,z,z')= \eta \left ( \frac{d}{\pi}\right )^2 \sum_{n=1}^{N}
\frac{4}{(2n-1) \pi} \\
&\sin \left ( \frac{2n-1}{2} \pi \frac{z}{L} \right ) \sin \left (
\frac{2n-1}{2} \pi \frac{z'}{L} \right ) \mathcal{M} \left (
\frac{r}{\xi_n}\right ) \, ,
\end{split} \eeq

\begin{figure}[h!tbp]
\centerline{\epsfig{file=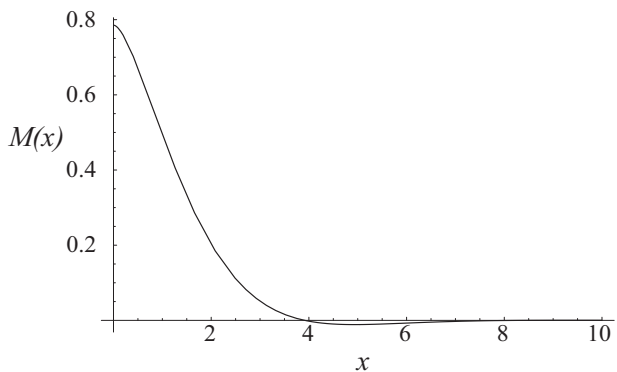,width=8cm}} \caption{The
$\mathcal{M}(x)$ function} \label{fig:Meijer}
\end{figure}

\noindent where the function $\mathcal{M}$ can be expressed in
terms of the Meijer G function\footnote{It can be represented in
\textsc{Mathematica} by the sequence~:
$\frac{1}{4}\mathrm{MeijerG} \left [ \left \{ \left \{ \right \},
\left \{ \right \} \right \}, \left \{ \left \{ 0, \frac{1}{2},
\frac{1}{2} \right \}, \left \{ 0 \right \} \right \}, \left(
\frac{x}{4} \right )^4 \right ]$}~: \beq \mathcal{M} \left ( x
\right ) = \int_{0}^{\infty} \dd q \, q \,\frac{\mathrm{J}_0 (q
x)}{1+q^4} = \frac{1}{4} \mathrm{G}_{0 \, 1}^{0 \, 4} \left (
\frac{x^4}{256} \left | _{\; 0,\frac{1}{2},\frac{1}{2},0} \right.
\right ) \label{Mofx} \eeq

A first observation is that the fluctuation amplitude $C(0,z,z)=
\left \langle \left | u(0,z) \right | ^2 \right \rangle $ is
\textit{always finite}; the physical reason is that the presence
of the substrate sets a lower boundary on the value of the wave
vector in the $z$ direction~: from equation (\ref{fnset}), $q_z
\geq \pi / 2L$, thus forbidding the soft mode with $q_z
\rightarrow 0$ and suppressing the Landau-Peierls instability.
Hence, there is no need for a lower cutoff in the integral in
equation (\ref{corrfun2}).

It is immediately obvious that $\mathcal{M} (0) = \pi / 4$, so the correlation function for
$r_{\bot}=0$ reduces to the simple formula~:
\beq
\label{corrfun0}
\begin{split}
C(0,z,z')= \eta \left ( \frac{d}{\pi}\right )^2
 \sum_{n=1}^{N} \frac{1}{2n-1} \\
 \sin \left ( \frac{2n-1}{2} \pi
\frac{z}{L} \right ) \sin \left ( \frac{2n-1}{2} \pi \frac{z'}{L}
\right )  \, , \end{split}
\eeq

\noindent which we shall use in determining the specular scattering
of the sample (subsection \ref{subsec:specular}). It is noteworthy
that this function varies as $C(0,z/d,z'/d) \sim \eta (d/\pi)^2$
for a fixed number of layers, so that any change in the
thermodynamic parameter $\eta$ only results in a scale factor.
This is very convenient for fitting the scattering spectrum of the
system (see section \ref{sec:scattering}), since the
time-consuming calculation of the correlation matrix between the
layers must only be performed once, and adjusting the $\eta$ and
$d$ parameters only changes a prefactor.

\begin{figure}[h!tbp]
\centerline{\epsfig{file=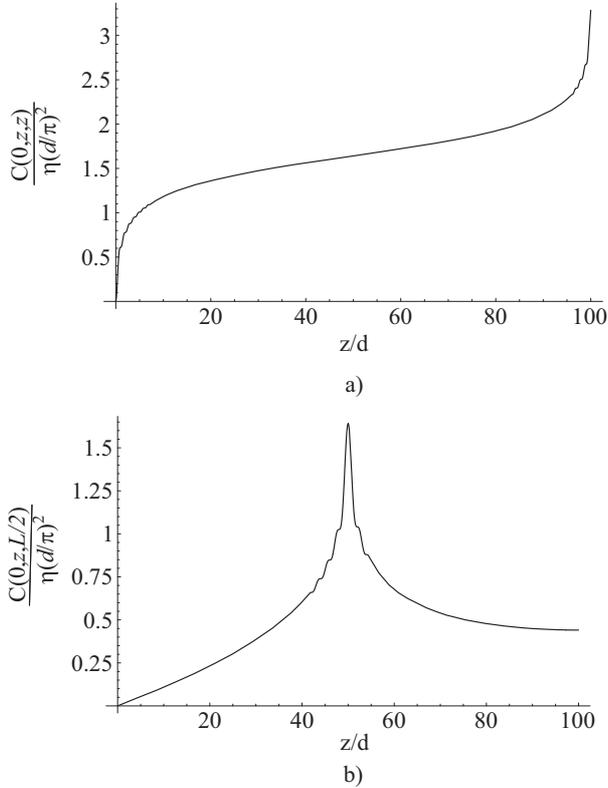,width=8cm}} \caption{The
correlation function (scaled by $\eta (d/\pi)^2$) in a 100 layer
film a) $C(0,z,z)$ b) $C(0,z,L/2)$.} \label{fig:Corr100_1}
\end{figure}

As an illustration, we present in Figure \ref{fig:Corr100_1} the
values of the correlation function in a stack of 100 bilayers; in
(a), $C(0,z,z)$ represents the fluctuation amplitude for each
bilayer, while in (b) $C(0,z,L/2)$ represents the correlation of
each bilayer with the one in the middle of the stack. Note the
sigmoidal shape of the function in the first case and the very
sharp peak in the second one.

Although for any finite number of layers $N$ the fluctuation
amplitude of the top layer $C(0,Nd,Nd)$ remains finite, it should
diverge in the bulk limit $N\rightarrow \infty$. From formula
(\ref{corrfun0}) one obtains~:
\beq
\begin{split}
\frac{C(0,Nd,Nd)}{\eta (d/\pi)^2} &= \sum_{n=1}^{N}\frac{1}{2n-1}\\
&= \frac{1}{2} \left [ \gamma + \ln 4 + \psi \left ( N -
\frac{1}{2} \right ) \right ]
\end{split}
\eeq

\noindent where $\gamma = 0.5772 \ldots$ is Euler's constant and
the digamma function $\psi (z)$ is the logarithmic derivative of
the gamma function, given by~: $\psi (z)=\Gamma ' (z)/\Gamma (z)$.
$C(0,Nd,Nd)$ diverges logarithmically, $\psi \left ( N - 1/2
\right )$ being indistinguishable from $\ln N$ as soon as $N>3$.

The same divergence spuriously appears even for a finite $L$ when the stack is treated as a completely
continuous medium, without limiting the number of $q_z$ modes to $N$ (formally, this amounts to letting
$N\rightarrow \infty$ and $d\rightarrow 0$ at fixed $L=Nd$). In this case, after summing over the $q_z$
modes, an artificial lower cutoff has to be introduced in the integral over $q_{\bot}$, as in reference
\cite{shalaginov93}. A comparison between our results and those obtained by this method is presented in the
Appendix.

\subsection{In-plane variation}
\label{subsec:inplane}

We shall now consider the $r$ variation of the correlation
function. At this point, it is convenient to introduce the
correlation of the height \textit{difference} $g(r,z,z')$ defined
by~:
\beq
\label{corrdiff}
\begin{split}
g(r,z,z')&= \left \langle \left ( u({\mathbf r}_{\bot},z)-
u({\mathbf 0},z') \right ) ^2 \right \rangle \\
&= C(0,z,z)+C(0,z',z')-2 C(r,z,z')\, .
\end{split}
\eeq

\noindent which has the advantage of remaining finite (for finite
values of $r$) even in unbound systems. We shall further write the
argument of the $\mathcal{M}$ function in equation
(\ref{corrfunfinal}) as~: $\ds \sqrt{\frac{(2n-1)\pi}{2N}}
\frac{r}{\xi}$, where $\xi=\sqrt{\lambda d}$ emphasizing that, for
a given number of layers, the $r$ variable in $C(r,z,z')$ and
$g(r,z,z')$ scales with the correlation length $\xi$. The physical
significance of this quantity can be seen as follows
\cite{helfrich78}~: for distances less than $\xi$, the layers
fluctuate independently, while for distances greater than $\xi$
the fluctuations are coherent from layer to layer. For an
unbounded medium (translation invariance along $z$) it was shown
\cite{lei95} that, for $z=z'$ the $g(r)$ function behaves as
\beq
g \left ( \frac{r}{\xi} \right )=\eta \left ( \frac{d}{\pi}
\right )^2 \frac{1}{2} \left ( \frac{r}{\xi} \right )^2 \left [
1-\gamma +\ln 2 -\ln \left ( \frac{r}{\xi} \right ) \right ]
\label{grsmall}
\eeq

\noindent if $r<\xi$ and as
\beq
g\left ( \frac{r}{\xi} \right )=\eta \left ( \frac{d}{\pi}
\right )^2 \left [ \ln \left ( \frac{r}{\xi} \right ) + \gamma
\right ] \label{grlarge}
\eeq

\noindent if $r>\xi$. In the limit $r \rightarrow \infty$, $g(r)$
diverges logarithmically, as it is well-known from the continuum
theory \cite{caille72}.

\begin{figure}[h!tbp]
\centerline{\epsfig{file=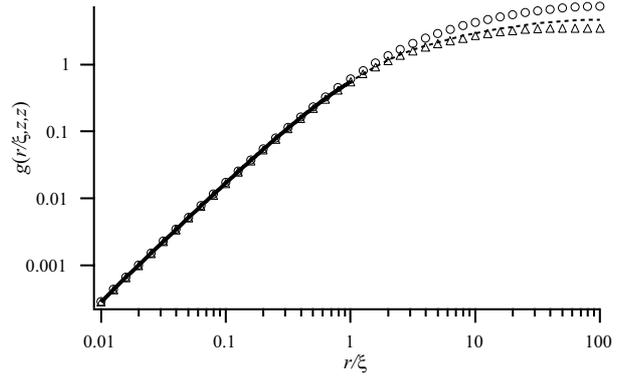,width=8cm}} \caption{The height
difference self-correlation function $g(r/\xi,z,z)$ scaled by
$\eta (d/\pi)^2$ in an 800 layer film, for $z=100d$ (triangles)
and $z=800d$ (circles). At small values of $r/\xi$, both curves
are well described by the asymptotic form (\ref{grsmall})
determined by Lei \cite{lei95} (solid line), with no adjustable
parameters. Also shown is the average value $g_{\textrm{avg}}(r)$
(dotted line).}
\label{fig:gofr}
\end{figure}

We present in Figure \ref{fig:gofr} the height difference
self-cor\-re\-la\-tion function $g(r/\xi,z,z)$, computed from
equations (\ref{corrdiff}) and (\ref{corrfunfinal}) for a stack of
800 bilayers.

In the low $r$ limit we find that, for all values of $z$, $g(r)$
is very well described by the asymptotic form (\ref{grsmall}). For
high values of $r/\xi$, the divergence (\ref{grlarge}) is replaced
by saturation to a value of $g(r \rightarrow \infty) = 2C(0)$.

As we shall see in subsection \ref{subsec:diffuse}, the
self-correlation function $g(r)$ can be extracted from the diffuse
scattering signal; however, when $g$ also depends on $z$ this
signal will be an average over the positions in the stack. For
comparison with the experimental data we calculate
$g_{\textrm{avg}}(r)$ (shown in Figure \ref{fig:gofr} as dotted
line) as an average over $g(r,z,z)$ for eight different values of
$z$, equally spaced from $100d$ to $800d$.

\section{Scattering}
\label{sec:scattering}

As discussed in the Introduction, notable differences appear
between the scattering signal from bulk smectic phases and from
solid-supported films. Even if the latter contain thousands of
bilayers, the clear separation between specular and diffuse
scattering shows that the Landau-Peierls effect is suppressed
\cite{salditt99}.

\subsection{Specular scattering}
\label{subsec:specular} In specular reflectivity studies, the
incidence and reflection angles are equal~: $\alpha _i = \alpha
_f$, corresponding to $q_{\bot}=0$. As we shall see below, in
this case only the correlation at zero in-plane distance (given by
formula (\ref{corrfun0})) is relevant. For simplicity and to
emphasize the discrete nature of the stack we shall use the
notation $C(0,nd,md) \equiv C(n,m)$.

The specular scattering factor of the lamellar stack (without
taking into account the substrate) $S(q_z,{\mathbf
q}_{\bot}={\mathbf 0})$ can be written as \cite{sinha88,lei95}~:
\beq
\label{specular} S(q_z)= \sum_{m,n=1}^N \cos [q_z d (m-n)]
\un{e}^{ -\frac{q_z^2}{2} (C(m,m)+C(n,n))}
\eeq

The spectra are corrected for the diffuse scattering by
subtraction of an offset scan \cite{salditt02}.

\begin{figure}[h!tbp]
\centerline{\epsfig{file=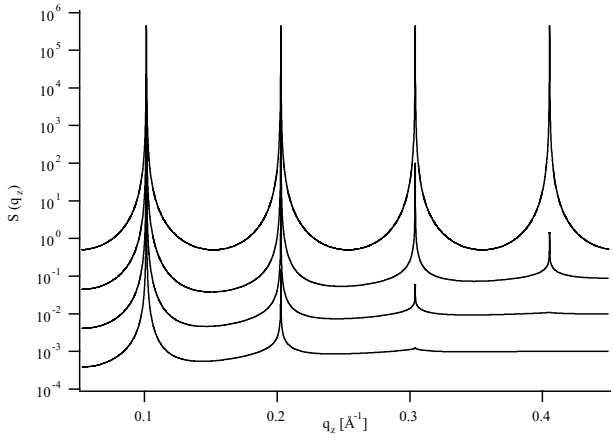,width=8cm}} \caption{Specular
structure factor for a stack of 1000 bilayers, with $d=62
\un{\AA}$, for different values of the Caill\'{e} parameter (from
top to bottom, $\eta=0$, $0.1$, $0.2$, and $0.3$). Curves
vertically shifted for clarity.} \label{fig:struct}
\end{figure}

\subsection{Accounting for partial dewetting}
\label{subsec:dewetting} Fully-hydrated lipid multilayers differ
from freely-sus\-pend\-ed smectic films in that their thickness is
not constant; partial dewetting leads to a variation in thickness
over the area of the sample \cite{perrinogallice02}. Notably, this
leads to an obliteration of the Kiessig fringes, due to
interference between the external interfaces of the system. We
take this effect into account by introducing a coverage rate for
each layer, $f(n)$, which is 1 at the substrate and decreases on
approaching the free surface. The structure factor then reads~:
\begin{eqnarray}
\label{eq:struct}
S(q_z)= \sum_{m,n=1}^N f(n)  f(m) \cdot \cos [q_z d
(m-n)]\\
  \cdot  \exp \left [ -\frac{q_z^2}{2} (C(m,m)+C(n,n)) \right ] \, .
\nonumber
\end{eqnarray}

For the coverage function we chose the convenient analytical form
$f(n)= \left [ 1-\left ( \frac{n}{N} \right ) ^{\alpha} \right ] ^2$,
where $\alpha$ is an empirical parameter controlling the degree of coverage.
This is a convenient method, but not a very precise one, insofar as the
fluctuation spectrum is still calculated for a fixed number of layers, $N$.
In the limit where the size of the domains with a given thickness is
larger than the X-ray coherence length, an alternative approach would be to
incoherently (no cross-terms between domains) average over a
distribution depending on the total layer number $P(N)$.

\subsection{Describing the reflectivity profile}
\label{subsec:reflect}

In order to model the reflectivity profile of our system, besides the structure factor of the stack one
needs the form factor of the bilayers. Furthermore, the presence of the substrate must be taken into
account. This is done using a semi-kinematic approximation, where the reflectivity of a rough interface is
expressed by the master-formula of reflectivity \cite{alsnielsen01}~:

\begin{equation}
\label{master} R(q_z)=R_{\un{F}}(q_z) \cdot \left | \frac{1}{\rho_{12}}\int_{0}^{\infty} \frac{\dd
\rho(z)}{\dd z} e^{iq_zz} \dd z \right |^2
\end{equation}

\noindent where $R_{\un{F}}$ denotes the Fresnel reflectivity of the sharp interface and $\rho(z)$ is the
intrinsic electron density profile whereas $\rho_{12}$ is the total step in electron density between the
two adjoining media (silicon and water, in our case).

For the electronic density profile of the bilayer we use a parameterization in terms of Fourier components
\cite{salditt02}, which has the advantage of describing the smooth profile of lipid bilayers using only a
few coefficients. For one bilayer we have~:

\beq \label{rhobl} \rho_{\un{bl}}(z) = \rho_{0}+ \sum_{k=1}^{N_{\un{comp}}} \rho_{k} v_k \cos \left (
\frac{2 \pi k z}{d} \right ) \eeq

\noindent where $\rho_{\un{bl}}(z)$ is defined between $-d/2$ and $d/2$, $N_{\un{comp}}$ is the total
number of Fourier components, $\rho_{k}$ is the amplitude of the $k$-th component and $v_k$ the associated
(complex) phase factor, which in our case can be shown to reduce to $\pm 1$ only, due to the mirror
symmetry of the bilayers. The bilayer form factor is given by the Fourier transform of $\rho_{\un{bl}}(z)$.

The total density profile is given by the density profiles of the $N$ bilayers (weighted by the coverage
factors $f$ described in subsection \ref{subsec:dewetting}) to which is added the profile of the substrate,
described by an error function of width the rms roughness of the \un{Si} wafer (this quantity can be
independently determined from the reflectivity of the blank wafer and is typically worth $8-10 \un{\AA}$
for all our measurements).

Effects related to the finite instrumental resolution and sample absorption are completely negligible in
thin samples and were not implemented.

\subsection{Comparison with experimental data}
\label{subsec:compexp}

\begin{figure*}[h!tbp]
\centerline{\epsfig{file=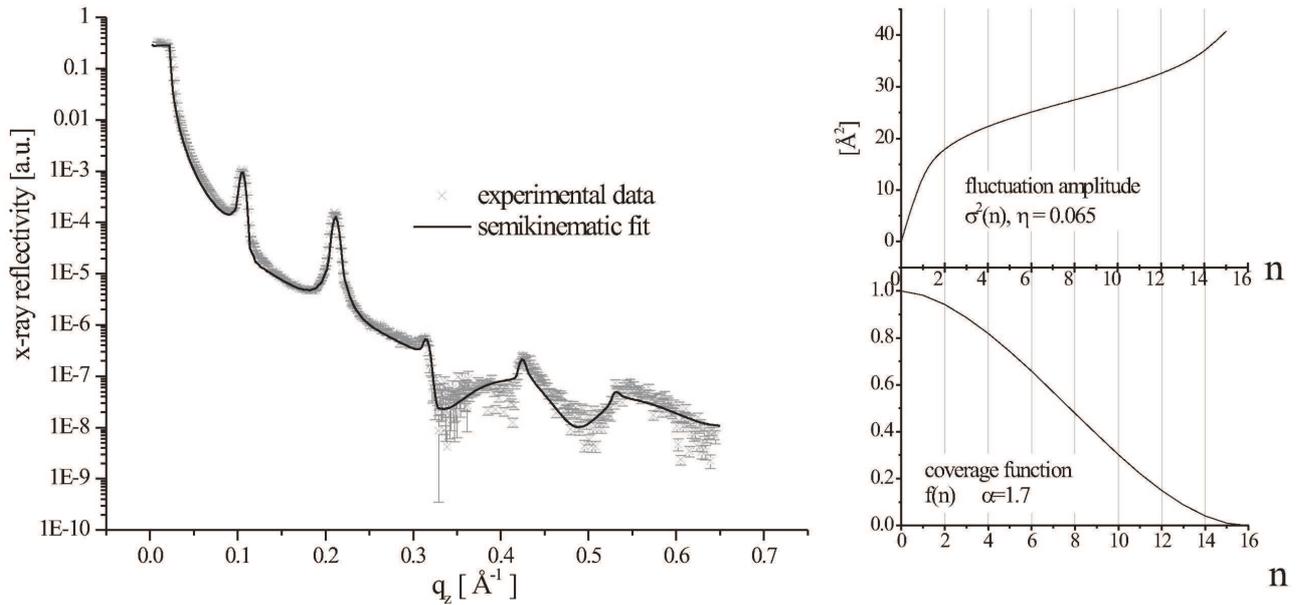, width=18cm}} \caption{X-ray
reflectivity curve of 16 DMPC bilayers on a Si substrate measured
under osmotic pressure (symbols) fitted with a full $q$-range semi-kinematic
model \cite{salditt02} using the structure factor (\ref{eq:struct}) (solid line). Inset~: (top) fluctuation
amplitude $\sigma ^2 (n)$ for a Caill\'{e} parameter $\eta=0.065$ obtained from the fit
and (bottom) coverage function $f(n)$ with $\alpha =1.7$ deduced from the fit.}
\label{peg36b}
\end{figure*}

Lipid films partially hydrated in humid atmosphere exhibit many well-defined Bragg peaks, indicative of
very low fluctuation amplitudes, so they are not a good testing ground for our model. On the other hand,
samples hydrated in excess pure water sometimes have \textit{too few} Bragg peaks. We thus chose to test
our model against spectra obtained on samples that are in excess solvent, but under an osmotic pressure
imposed by PEG (polyethylene glycol) solutions \cite{mennicke03}.
 Fig. \ref{peg36b} shows the X-ray reflectivity of
16 1,2-dimyristoyl-sn-glycero-3-phosphocholine (DMPC) bilayers on a silicon substrate measured in an
aqueous PEG solution of 3.6 wt. \% concentration. The temperature was maintained at $40 \dgr$, at which the
lipids are in the fluid $L_{\alpha}$ phase. The continuous line shows a full $q$-range fit to the data
using the structure factor (\ref{eq:struct}). From the position of the Bragg peaks one can determine a
periodicity of $59.5 \un{\AA}$. The suppression of the higher-order Bragg peaks clearly indicates the
influence of the fluctuations. The fit yielded the parameters: $\eta=0.065$ for the Caill\'{e} parameter
and $\alpha=1.7$ for the coverage exponent in \ref{eq:struct}. The corresponding fluctuation amplitudes
$\sigma^2(n)=C(n,n)$ and coverage function $f(n)$ are also shown. Further details on the results of these
measurements will be presented elsewhere \cite{mennicke03}.

{\bf Experimental details~:} The curve in Fig. \ref{peg36b} was
measured at the bending magnet beamline D4 at the HASYLAB/Desy in Hamburg, Germany. After
the beam passed a Rh-Mirror to reduce high energy flux, a photon energy of $19.92 \un{keV}$
was chosen using a single Si(111) crystal monochromator. The beam was collimated with several
motorized vertical and horizontal slits. The reflected intensity
was measured with a fast scintillation counter (Cyberstar, Oxford).

{\bf Sample preparation:} The lipid 1,2-di\-my\-ris\-to\-yl-sn-gly\-ce\-ro-3-phosphocholine (DMPC) was
bought from A\-van\-ti and used without further purification. 16 bilayers were prepared on a commercial
silicon substrate using the spin-coating method as described in \cite{mennicke02}. The substrate was cut to
a size of $15 \times 25 \un{mm^2}$ and carefully cleaned in an ultrasonic bath with methanol for ten
minutes and subsequently rinsed with methanol and ultrapure water (Milli-Q, Millipore). The lipid was
dissolved in chloroform at a concentration of $10 \un{mg/ml}$. An amount of $100 \un{\mu l}$ of the
solution was pipetted onto the cleaned and dried substrate which was then accelerated to $3000 \un{rpm}$
using a spin-coater. After rotation for 30 seconds, the sample was exposed to high vacuum over night in
order to remove all remaining traces of solvent. The sample was refrigerated until the measurement. For the
measurement the sample was mounted in a stainless steel chamber with kapton windows which can be filled and
in situ flushed with fluids such as water or polymer solutions. PEG of molar weight 20.000 was bought from
Fluka and used without further purification. A concentration of 3.6 wt. \% corresponds to an osmotic
pressure of about $10^{4} \un{Pa}$ \cite{rand}.

\subsection{Diffuse scattering}
\label{subsec:diffuse}

It is well-known that the diffuse scattering signal from a surface can be expressed in terms of the height
difference self-cor\-re\-la\-tion function $g(r)$ of the surface \cite{sinha88}. For a multilayer system,
one must also take into account the cross-correlation function $g(r,i,j)$ \cite{sinha94} and the
calculations become much more complicated. However, it can be shown that by integrating the diffuse signal
in $q_z$ over a Brillouin zone, the cross-correlation terms $i\neq j$ cancel and one is left with a curve
corresponding to a transform of an average self-cor\-re\-la\-tion function \cite{salditt94}.

\begin{figure}[h!tbp]
\centerline{\epsfig{file=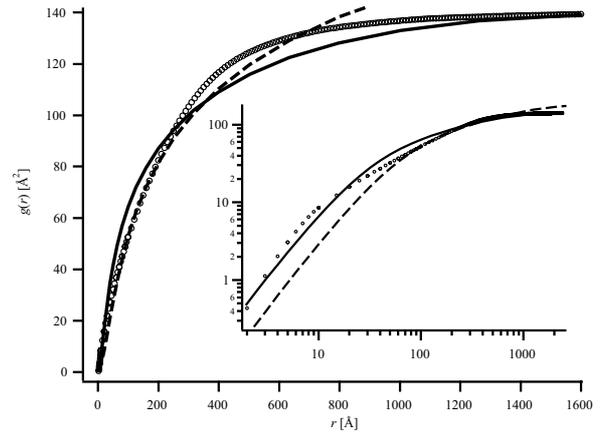,width=8cm}} \caption{Experimental values for the self-correlation
function $g_{\textrm{exp}}(r)$ (open dots) and $g_{\textrm{avg}}(r)$ for two sets of fitting parameters~:
$\eta =0.071$ and $\xi =20 \un{\AA}$ (solid line) and $\eta =0.09$ and $\xi =40 \un{\AA}$ (dashed line).
Inset~: the same graph in log-log scaling.} \label{fig:fit_gofr}
\end{figure}

We compare our model to recent experimental data obtained on stacks of 800 fully-hydrated DMPC bilayers,
where the experimental self-cor\-re\-la\-tion function $g_{\textrm{exp}}(r)$ was obtained by inverting the
integrated diffuse scattering \cite{salditt03}. In Figure \ref{fig:fit_gofr} we present
$g_{\textrm{exp}}(r)$ (open dots), as well as the $g_{\textrm{avg}}(r)$ function defined in subsection
\ref{subsec:inplane}, with two different sets of fitting parameters $\eta$ and $\xi$.

The theoretical curve can be scaled to the experimental data, either by adjusting both parameters, yielding
$\eta =0.071$ for the Caill\'{e} parameter and $\xi = \sqrt{\lambda d} = 20 \un{\AA}$ (solid line in Fig.
\ref{fig:fit_gofr}) or with a fixed parameter $\xi = 40 \un{\AA}$ (from the analysis of the half width of
the diffuse Bragg sheets \cite{salditt03}), and an open parameter $\eta$ which is adjusted to $\eta\simeq
0.09$. Note that the present expression for $g(r)$ can give better account of the experimental data than
the bulk correlation function used before \cite{lei95,salditt03}. In particular, it reproduces the observed
saturation regime at high $r$, while the bulk correlation function diverges logarithmically. However,
systematic discrepancies between data and theory should not be overlooked. We note that the functional form
of $g(r)$ for small $r\ll \xi_N$ is not well captured, in particular it can not explain the exponent
$g(r)\propto r^{0.7}$ of the algebraic regime in the data. The possible reasons  for this discrepancy can
be manifold: contributions of non-bending modes to the diffuse scattering, a length scale dependent bending
rigidity $\kappa$, residual tension in the bilayers due to edge effects, or nonlinearities in the
Hamiltonian. This needs to be investigated in future studies. Finally, a more detailed treatment of the
$z$-dependence is needed, since the experimental analysis determines a correlation function which is
averaged over the scattering volume.

\section{Conclusion}

The presence of a substrate dramatically changes the thermal
fluctuations in lipid multilayers. Most noticeably, the
Landau-Peierls instability is suppressed. We present a theoretical model
taking into account this feature and show that it describes very
well the experimental reflectivity data. Reasonable agreement is also
obtained for the diffuse scattering. Our result is a first step
towards a unitary interpretation of the global (specular and
diffuse) scattering signal of solid-supported lipid multilayers.

\appendix

\section*{Appendix. Comparison with previous results}
\label{app:comparison}

\begin{figure}[h!tbp]
\centerline{\epsfig{file=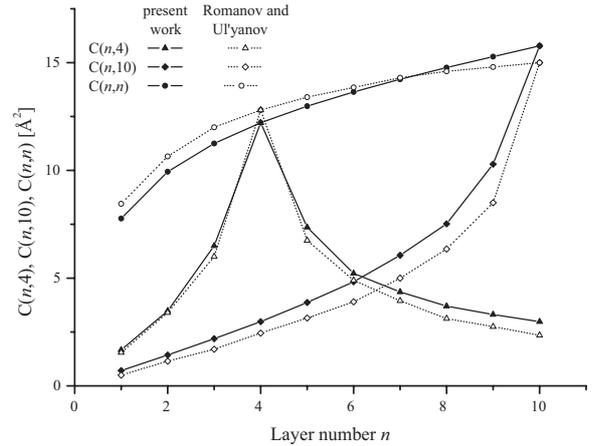,width=8cm}} \caption{Comparison
between the results of the present method, for $\gamma =0$ (in
full symbols and solid line) and the values obtained by Romanov
and Ul'yanov \cite{romanov02} for $\gamma =30 \un{mN/m}$ (open
symbols, dotted line) for the correlation function in a film
comprising 21 layers (with $d=30 \un{\AA}$ and $\eta=0.14$).
Values only shown for the bottom half of the stack.}
\label{fig:Comp}
\end{figure}

In order to assess the validity of our method, we compared our
results with those obtained by Romanov and Ul'yanov by a rigorous
treatment of the discrete model \cite{romanov02}. However, they
only show data for smectic films with a rather high surface
tension ($\gamma =30 \un{mN/m}$). Thus, we chose their thickest
film (21 layers) and only compared the values of $C$ for the
bottom half (close to the substrate), where the surface tension
plays a lesser role. As shown in Figure \ref{fig:Comp}, the
agreement is quite good.

We also compared our results to those obtained by Shalaginov and
Romanov \cite{shalaginov93}, who also used a continuous model, but
without restricting the number of modes. For a solid substrate and
a vanishing surface tension at the top, their equations (18a) and
(22) lead to~:
\beq
\label{shalaginov}
\begin{split}
&C(m,n)= \frac{k_B T}{8 \pi\sqrt{BK}} \int_{0}^{\frac{2\pi}
{a_{0}}} \frac{\dd x}{x \cosh (x N)} \\
&[ \sinh [x(n+m-N)] + \sinh (x N) \cosh [x(n-m)] \\
&-\cosh (x N) \sinh (x \left | n-m \right | ) ]
\end{split}
\eeq

\begin{figure}[h!tbp]
\centerline{\epsfig{file=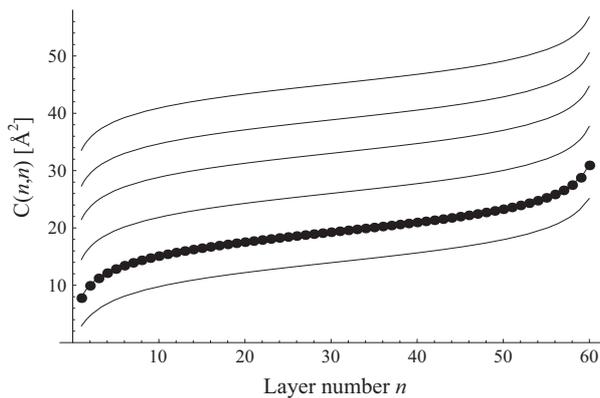,width=8cm}}
\caption{Comparison between the results of the present method
(solid dots) and the values obtained by the method of Shalaginov
and Romanov \cite{shalaginov93} (lines), for a film comprising 61
layers. From top to bottom, the values of the cutoff parameter
$a_0$ in equation (\ref{shalaginov}) are~: 1.5, 4, 10, 30, 85, and
200 \AA.} \label{fig:Shalaginov}
\end{figure}

In Figure \ref{fig:Shalaginov} we present our results for $C(n,n)$
in a film with 61 layers, with $d=30 \un{\AA}$ and $\eta=0.14$
(solid dots) as well as the values obtained using the formula
(\ref{shalaginov}), for different values of the cutoff parameter
$a_0$ (from top to bottom, 1.5, 4, 10, 30, 85, and 200 \AA).
Clearly, the fluctuation amplitude is very sensitive to the value
of $a_0$, with an approximately logarithmic dependence, and gives
the same results as our method for $a_0=85 \un{\AA}$. We remind
that neither in our method, nor in the one of Romanov and Ul'yanov
is there any need for a cutoff, due to mode number limitation (see
the discussion in subsection \ref{subsec:gammazero}).

\begin{acknowledgement}
D. C. has been supported by a Marie Curie Fellowship of the European
Community programme \textit{Improving the Human Research Potential} under contract number HPMF-CT-2002-01903.
\end{acknowledgement}


\begin{thebibliography}{}

\bibitem{katsaras00} J. Katsaras and V. A. Raghunathan, \textit{Aligned lipid-water
systems}, in \textit{Lipid bilayers~: Structure and Interactions},
edited by J. Katsaras and T. Gutberlet (Springer, 2000).

\bibitem{salditt02} T. Salditt, C. Li, A. Spaar, and U. Mennicke, Europhys. J. E {\bf 7}, 105-116 (2002).

\bibitem{lyatskaya00} Y. Lyatskaya, Y. Liu, S. Tristram-Nagle, J. Katsaras, and J. F. Nagle, Phys. Rev. E {\bf 63}, 011907 (2000).

\bibitem{bechinger97} B. Bechinger J. Membrane Biol. {\bf 156}, 197-211 (1997).

\bibitem{caille72} A. Caill\'{e}, C. R. Acad. Sci. Paris, S\'{e}r. B {\bf 274}, 891-893 (1972).

\bibitem{alsnielsen80} J. Als-Nielsen, J. D. Litster, D. Birgenau, M. Kaplan, C. R. Safinya,
A. Lindgard-Andersen, and S. Mathiesen, Phys. Rev. B {\bf 22}, 312-320 (1980).

\bibitem{safinya86} C. R. Safinya, D. Roux, G. S. Smith, S. K. Sinha, P. Dimon, N. A. Clark, and A.-M. Bellocq, Phys. Rev. Lett. {\bf 57}, 2718-2721 (1986).

\bibitem{safinya89} C. R. Safinya, D. Roux, G. S. Smith, Phys. Rev. Lett. {\bf 62}, 1134-1137 (1989).

\bibitem{nallet93} F. Nallet, R. Laversanne, and D. Roux, J. Phys II (France) {\bf 3}, 487-502 (1993).

\bibitem{zhang94} R. Zhang, R. M. Sutter, and J. F. Nagle, Phys. Rev. E {\bf 50}, 5047-5060 (1994).

\bibitem{nagle96} J. F. Nagle, R. Zhang, S, Tristram-Nagle, W. Sun, H. I. Petrache and R. M. Suter, Biophys. J. {\bf 70}, 1419-1431 (1996).

\bibitem{pabst00} G. Pabst, M. Rappolt, H. Amenitsch, and P. Laggner, Phys. Rev. E {\bf 62}, 4000-4009 (2000).

\bibitem{holyst91} R. Ho{\l}yst, Phys. Rev. A {\bf 44}, 3692-3709 (1991).

\bibitem{shalaginov93} A. N. Shalaginov and V. P. Romanov, Phys. Rev. E {\bf 48}, 1073-1083 (1993).

\bibitem{lei95} N. Lei, C. R. Safinya and R. F. Bruinsma, J. Phys. II France {\bf 5}, 1155-1163 (1995); N. Lei, Ph. D. Thesis, Physics Department, Rutgers, The State University of New Jersey (1993).

\bibitem{mol96} E. A. L. Mol, J. D. Shindler, A. N. Shalaginov and W. H. de Jeu, Phys. Rev. E {\bf 54}, 536-549 (1996).

\bibitem{oswald02} P. Oswald and P. Pieranski, \textit{The Liquid Crystals~: Concepts and Physical Properties Illustrated by Experiments} (in French), (CPI, Paris, 2002).

\bibitem{deboer99} D. K. G. de Boer, Phys. Rev. E {\bf 59}, 1880-1886 (1999).

\bibitem{romanov02} V. P. Romanov and S. V. Ul'yanov, Phys. Rev. E {\bf 66}, 061701 (2002).

\bibitem{poniewierski93} A. Poniewierski and R. Ho{\l}yst, Phys. Rev. B {\bf 47}, 9840-9843 (1993).

\bibitem{safran99} S. A. Safran, Adv. Phys. {\bf 48}, 395-448 (1999).

\bibitem{helfrich78} W. Helfrich, Z. Naturforsch. A {\bf 33}, 305-315 (1978).

\bibitem{salditt99} T. Salditt, C. M\"{u}nster, J. Lu, M. Vogel, W. Fenzl, and A. Souvorov, Phys. Rev. E {\bf 60}, 7285-7289 (1999).

\bibitem{perrinogallice02} L. Perino-Gallice, G. Fragneto, U. Mennicke, T. Salditt, and F. Rieutord, Europhys. J. E {\bf 8}, 275-282 (2002).

\bibitem{alsnielsen01} J. Als-Nielsen and D. McMorrow, \textit{Elements of Modern X-Ray Physics}, (Wiley, Chichester, 2001).

\bibitem{mennicke03} U. Mennicke, D. Constantin, and T. Salditt, in preparation.

\bibitem{mennicke02} U. Mennicke and T. Salditt, Langmuir {\bf 18}, 8172-8177 (2002).

\bibitem{rand} The data was obtained from the web site of the Membrane Biophysics Laboratory at the Brock
University in Canada~: http://aqueous.labs.brocku.ca/osfile.html. The value for the osmotic pressure
induced by a PEG 20000 solution at 3.6 wt. \% is only available at $20 \dgr$ as $1.4 \, 10^{4} \un{Pa}$. At
$40 \dgr$, temperature at which the experiments were performed, we estimate that the pressure is lower by
about 10 to 20 \%, by using the temperature dependence of PEG 8000 solutions (available on the same site).

\bibitem{sinha88} S. K. Sinha, E. B. Sirota, S. Garoff, and H. B. Stanley, Phys. Rev. B {\bf 38}, 2297-2312 (1988).

\bibitem{sinha94} S. K. Sinha, J. Phys. (France) III {\bf 4}, 1543-1557 (1994).

\bibitem{salditt94} T. Salditt, T. H. Metzger, and J. Peisl, Phys. Rev. Lett. {\bf 73}, 2228-2231 (1994), T. Salditt, T.H. Metzger, Ch. Brandt, U. Klemradt, and J. Peisl, Phys. Rev. B {\bf51}, 5617-5627 (1995).

\bibitem{salditt03} T. Salditt, M. Vogel, and W. Fenzl, Phys. Rev. Lett. {\bf 90}, 178101 (2003).

\end{thebibliography}
\end{document}